\documentclass[prb,twocolumn,superscriptaddress,showpacs,%
preprintnumbers,amsmath,amssymb]{revtex4}
\usepackage{amsmath}
\usepackage{graphicx}
\usepackage{dcolumn}

\begin{document}

\title{(Giant) Vortex - (anti) vortex interaction in bulk superconductors: The Ginzburg-Landau theory}

\author{Andrey Chaves}\email{andrey@fisica.ufc.br}
\affiliation{Department of Physics, University of Antwerp,
Groenenborgerlaan 171, B-2020 Antwerp,
Belgium}\affiliation{Departamento de F\'isica, Universidade
Federal do Cear\'a, Caixa Postal 6030, Campus do Pici, 60455-900
Fortaleza, Cear\'a, Brazil}
\author{F. M. Peeters}\affiliation{Department of Physics, University of
Antwerp, Groenenborgerlaan 171, B-2020 Antwerp,
Belgium}\affiliation{Departamento de F\'isica, Universidade
Federal do Cear\'a, Caixa Postal 6030, Campus do Pici, 60455-900
Fortaleza, Cear\'a, Brazil}
\author{G. A. Farias}\email{gil@fisica.ufc.br}
\affiliation{Departamento de F\'isica, Universidade Federal do
Cear\'a, Caixa Postal 6030, Campus do Pici, 60455-900 Fortaleza,
Cear\'a, Brazil}
\author{M. Milo\v{s}evi\'c}\affiliation{Department of Physics, University of
Antwerp, Groenenborgerlaan 171, B-2020 Antwerp, Belgium}

\date{ \today }

\begin{abstract}
The vortex-vortex interaction potential in bulk superconductors is
calculated within the Ginzburg-Landau (GL) theory and is obtained
from a numerical solution of a set of two coupled non-linear GL
differential equations for the vector potential and the
superconducting order parameter, where the merger of vortices into
a giant vortex is allowed. Further, the interaction potentials
between a vortex and a giant vortex and between a vortex and an
antivortex are obtained for both type-I and type-II
superconductors. Our numerical results agree asymptotically with
the analytical expressions for large inter-vortex separations
which are available in the literature. We propose new empirical
expressions valid over the full interaction range, which are
fitted to our numerical data for different values of the GL
parameter.

\end{abstract}


\maketitle

\section{Introduction}

The interaction potential between vortices has been an important
study subject for many years. In 1971, Kramer \cite{Kramer} used
the asymptotic behavior of the vortex fields for large distances
in the Abelian Higgs model to obtain an analytical expression for
the vortex-vortex interaction potential, which is given by
modified Bessel functions. This potential is attractive
(repulsive) for type-I (type-II) systems, i.e., for a
superconductor Ginzburg-Landau parameter with $\kappa =
\lambda/\xi < 1/\sqrt{2}$ ($> 1/\sqrt{2}$), where $\lambda$ is the
penetration depth and $\xi$ is the coherence length. Moreover,
this expression leads to a constant interaction energy as a
function of the separation between vortices for the critical value
$\kappa = 1/\sqrt{2}$ (also called the Bogomol'nyi point),
implying that vortices do not interact in this regime. A detailed
analysis of the vortex-vortex (V-V) interaction was given later by
Jacobs and Rebbi, \cite{Jacobs} who constructed a variational
function describing two separate vortices and obtained the
variational parameters by minimizing the free energy. This
variational function was able to model: i) the deformation of the
vortex core when the vortices are brought close to each other, and
ii) the formation of a giant vortex \cite{Schweigert, Kanda,
Golubovic} when vortices are superimposed on each other.

Thereafter, several works studied different aspects of the
interaction between vortices in superconductors. For example,
Brandt \cite{Brandt} used the asymptotic expression for the
interaction potential to study the elastic properties of flux-line
lattices in type-II superconductors. Speight \cite{Speight}
derived the V-V interaction from a linear field theory described
by a Lagrangian of two singular point sources placed at the vortex
centers. MacKenzie \emph{et al.} \cite{MacKenzie} used the linear
field theory proposed by Speight to obtain the interaction between
separated strings in a model with two order parameters, which may
be relevant for superconducting cosmic strings, \cite{Bettencourt}
the SO(5) model of high-temperature superconductivity and solitons
in nonlinear optics. Eventually, all these analytical
approximations lead to expressions which are the same, or at least
very similar, to the one derived by Kramer for large separation
between vortices. Other models were presented by Mohamed \emph{et
al.} \cite{Mohamed}, who used a perturbative approach to calculate
the V-V interaction for superconductors with $\kappa \approx
1/\sqrt{2}$ when the Ginzburg-Landau (GL) theory is extended to
low temperatures, and by Hern\'andez and L\'opez, \cite{Hernandez}
who used a variational approach based on the Clem trial function
\cite{Clem} to calculate the force between vortices. Auzzi
\emph{et al.} \cite{Auzzi} showed recently that for non-Abelian
vortex interactions, there are two other regimes besides the
well-known type-I and type-II: depending on the relative
orientation, the interaction potential can show attractive and
repulsive regions for the same system. Similar behavior can be
achieved in a two-band superconductor. \cite{Babaev, Roeland0}

In the present work, we solve numerically the set of
Ginzburg-Landau (GL) differential equations for two vortices fixed
at a certain distance from each other. We are able to recover the
V-V potential obtained by Jacobs and Rebbi \cite{Jacobs} and the
asymptotic behaviors predicted by Kramer \cite{Kramer}. The
results are generalized to the case of vortex-giant vortex (V-GV)
and vortex-antivortex (V-AV) interactions. We parameterize all
obtained forces for future use in e. g. molecular dynamics
simulations of the vortex matter.

The remainder of this paper is organized as follows: in Sec. II,
we review the asymptotic behavior of the magnetic field and order
parameter of the single static vortex in the GL theory and expand
the analytical expression suggested by Kramer to study the
interaction between a single vortex and another vortex, an anti
vortex or a giant vortex, in the limit of large separation between
them. In Sec. III we discuss the set of coupled non-linear
differential equations that describe these interactions, which are
valid for arbitrary separation between vortices, and thus, also
for small distances where deformations of the interacting vortex
cores are important. This set of equations is solved numerically
without any approximations for arbitrary values of $\kappa$, and
the results for the V-V, V-GV and V-AV interactions are shown in
Sec. IV and compared to the analytical expressions obtained in
Sec. II. For each of these cases, a fitting function for the
interaction is proposed and the fitting parameters are given. Our
results are summarized in Sec. V.

\section{Asymptotics of the inter-vortex potential}

Let us start with the expression for the free energy in the GL
theory, or equivalently, the potential in the Abelian Higgs model
\cite{Jacobs, Babaev}:
\begin{equation}
E = \int \mathcal{F} d\vec{\textsf{r}}, \label{energy}
\end{equation}
where the functional $\mathcal{F}$ is given by
\begin{equation}
\mathcal{F} = \frac{\hbar
^2}{4m}\left|\left(\vec{\nabla}-i\frac{2e}{\hbar
c}\vec{\textsf{A}}\right)\psi\right|^2 +
\frac{1}{8\pi}|\vec{\nabla} \times \vec{\textsf{A}}|^2 +
\frac{\beta}{2}|\psi|^4 - \alpha|\psi|^2. \label{lagrangean}
\end{equation}
In this expression, $\psi$ is the order parameter (a complex
scalar field), $\vec{\textsf{A}}$ is the electromagnetic vector
(gauge) potential and $\alpha$ and $\beta$ are phenomenological
parameters, which are related to the two characteristic lengths in
a superconductor: the coherence length $\xi = \hbar/\sqrt{4 m
\alpha}$ and the penetration depth $\lambda =
c/e(\sqrt{m\beta/8\pi\alpha})$. It is convenient to define $\mu =
\sqrt{2}\kappa$, so that when $\mu < 1$ ($\mu > 1$) we are in the
type-I (type-II) regime ($\kappa = 1/\sqrt{2}$ leads to $\mu =
1$). \cite{Jacobs} If we define $\vec{A} = (2e/\hbar c)
\vec{\textsf{A}}$ and $\Psi = \sqrt{\beta/\alpha}\psi/\lambda$,
the expression for the energy functional can be rewritten in
dimensionless units as
\begin{equation}
F = \frac{1}{2}|(\vec{\nabla}-i\vec{A})\Psi|^2 +
\frac{1}{2}|\vec{\nabla} \times \vec{A}|^2 +
\frac{\mu^2}{8}(1-|\Psi|^2)^2, \label{lagrangeanUnits}
\end{equation}
where the distances are scaled by the magnetic field penetration
depth $\lambda$, energy by $E_0 = \beta\big/2\alpha^2\xi^2$ and
the force by $\Omega_0 = \beta\big/2\alpha^2\xi^2\lambda$.

The lowest energy configuration of the system is found by
minimizing $E$ with respect to $\Psi$ and the vector potential
$\vec{A}$. The standard way of minimizing a functional is by the
Euler-Lagrange equations, which in the case of Eq. (\ref{energy})
are
\begin{subequations}
\begin{equation}
\frac{\partial F}{\partial \Psi} -
\sum_{i}\frac{\partial}{\partial x_i}\frac{\partial F}{\partial
(\frac{\partial \Psi}{\partial x_i})} = 0 \label{EL1}
\end{equation}
and
\begin{equation}
\frac{\partial F}{\partial A_j} - \sum_{i}\frac{\partial}{\partial
x_i}\frac{\partial F}{\partial (\frac{\partial A_j}{\partial
x_i})} = 0. \label{EL2}
\end{equation}
\end{subequations}
Equations (\ref{EL1}) and (\ref{EL2}) result in the well-known
Ginzburg-Landau equations. \cite{Tilley}

To find the V-V interaction potential, one has to control the
localization and the winding number (also called vorticity) of the
vortices. As mentioned earlier, previous theoretical works propose
a way to fix the vortices and obtain the interaction potential,
based on four steps: i) fixing a circular phase change of 2$\pi$
around each vortex, ii) finding the asymptotic behaviors of the
vector potential and the amplitude far from each of the vortices
\cite{Kramer} or, equivalently, solving numerically the
differential equations for these variables, \cite{Jacobs} iii)
constructing a superposition ansatz for $\Psi$ and $\vec{A}$ which
describes the double vortex structure, and iv) using the latter
ansatz in Eqs. (\ref{energy}) and (\ref{lagrangeanUnits}) to
obtain the energy for a given separation between vortices.

If we have the asymptotics of the vortices, obtained in step ii),
the integral in Eq. (\ref{energy}) can be solved analytically,
giving an analytical expression for the energy as a function of
the separation between vortices. In order to find the analytical
expression for the asymptotic interaction potential between a
vortex and another vortex, an antivortex or a giant-vortex, we
start from the sequence listed above: a circular phase is fixed
around each vortex by assuming $\Psi(r, \theta) = f(r)e^{i n
\theta}$, where $(r, \theta)$ are polar coordinates with the
origin in the center of the vortex, $n$ is its winding number and
$f(r)$ is the amplitude of its order parameter, which is assumed
to be circularly symmetric around the vortex center. Considering
the gauge $\vec{A} = na(r)\widehat{\theta}/r$, the Euler-Lagrange
equations (\ref{EL1}) and (\ref{EL2}) for a single vortex read
\begin{subequations}
\begin{equation}
\frac{d^2f}{dr^2} + \frac{1}{r}\frac{df}{dr}
-\frac{n^2(a-1)^2}{r^2}f - \frac{\mu^2}{2}f(f^2-1) = 0,
\label{ELcirc1}
\end{equation}
and
\begin{equation}
\frac{d^2a}{dr^2} - \frac{1}{r}\frac{da}{dr} - (a-1)f^2 = 0,
\label{ELcirc2}
\end{equation}
\end{subequations}
with $f(\infty) = a(\infty) = 1$. As is well known \cite{Jacobs},
the winding number $n$ also determines the number of zeros of the
vortex field $f(r)$ and, because of the circular symmetry, these
zeros must be degenerate at $r = 0$. Substituting the auxiliary
functions $\sigma(r)=f(r)-1$ and $Q(r)=a(r)-1$ in Eqs.
(\ref{ELcirc1}) and (\ref{ELcirc2}), we can suppress the high
order terms in the remaining differential equations when $r
\rightarrow \infty$, as $\sigma(\infty) = Q(\infty) = 0$, which
leads to the following equations, valid in the asymptotic limit:
\begin{subequations}
\begin{equation}
\left[\frac{d^2\sigma}{d(\mu r)^2} + \frac{1}{\mu
r}\frac{d\sigma}{d(\mu r)}-\sigma\right] = 0, \label{ELAsymp1}
\end{equation}
and
\begin{equation}
\frac{d^2}{dr^2}\left(\frac{Q}{r}\right) +
\frac{1}{r}\frac{d}{dr}\left(\frac{Q}{r}\right)-\left(1+\frac{1}{r^2}\right)\left(\frac{Q}{r}\right)
= 0. \label{ELAsymp2}
\end{equation}
\end{subequations}
Equations (\ref{ELAsymp1}) and (\ref{ELAsymp2}) are easily
identified as modified Bessel equations and their solutions are
$\sigma(r) = \gamma_1K_0(\mu r)$ and $Q(r) = \gamma_2rK_1(r)$
where $\gamma_1$ and $\gamma_2$ are coefficients to be determined.
For example, after solving Eqs. (\ref{ELcirc1}) and
(\ref{ELcirc2}) numerically, one can obtain these coefficients by
fitting $f(r) = 1+\gamma_1K_0(\mu r)$ and $\overrightarrow{A} =
n(1+\gamma_2rK_1(r))\widehat{\theta}/r$ to the results obtained by
the numerical procedure.

Several different procedures can be followed to extract the
analytical expression for the interaction potential from these
asymptotic functions. As an example, Bettencourt and Rivers
\cite{Bettencourt} suggested that one can substitute the
superposition ansatz $\overline{\Psi}(r,r_1,r_2) =
\Psi(|r-r_1|)\Psi(|r-r_2|)$ and $\overline{A_{\theta}}(r,r_1,r_2)
= A_{\theta}(|r-r_1|)+A_{\theta}(|r-r_2|)$, for vortices centered
at $\vec{r_1}$ and $\vec{r_2}$, in the energy functional in Eq.
(\ref{energy}) and keep only the terms that are linear in the
fields for each vortex. After some calculations described in more
detail in Ref. \cite{Bettencourt}, one obtains
\begin{equation}
E_{int}(d) =
2\left[n_1n_2\gamma_2^{(1)}\gamma_2^{(2)}K_0(d)-\gamma_1^{(1)}\gamma_1^{(2)}K_0(\mu
d)\right], \label{Eint}
\end{equation}
where $d$ is the separation between vortices, $n_i$ is the winding
number and $\gamma_1^{(i)}$ and $\gamma_2^{(i)}$ are the fitting
coefficients for the vortex $i$ in position $\vec{r_i}$. We point
out that in the paper by Bettencourt and Rivers, the expression
for $E_{int}$ is slightly different from Eq. (\ref{Eint}) because
they assumed that $\gamma_1^{(i)} = |n_i|\gamma_2^{(i)}$, which is
valid only in the critical coupling regime for $\mu = 1$, as
stressed by Speight \cite{Speight} and verified by Bogomol'nyi
equations for this regime. The same expression for $E_{int}(d)$
was found by Kramer by a perturbational approach \cite{Kramer} and
can also be obtained by considering point sources in a linearized
field theory \cite{Speight, MacKenzie}.

It can be easily seen that the potential in Eq. (\ref{Eint}) is
consistent with the fact that for $\mu < 1$ ($> 1$), the V-V
interaction is attractive (repulsive) for large separation,
leading to a type-I (type-II) superconducting behavior. This
statement is also valid for vortex-giant vortex interactions, as
Eq. (\ref{Eint}) still holds for this case, where the interaction
potential is obtained just by setting $n_1 = 1$ and $n_2 > 1$ and
finding the fitting coefficients $\gamma_1^{(i)}$ and
$\gamma_2^{(i)}$ for this case. However, for a V-AV interaction,
$n_1n_2$ is always negative, leading to an \emph{attractive}
potential $E_{int}(d)$ for \emph{any value} of $\kappa$.
Furthermore, Eq. (\ref{Eint}) shows that the interaction between
an antivortex and a giant vortex is always attractive as well.
This can be understood through a heuristic argument: when a vortex
and an antivortex are far from each other, the energy of the
system is non-zero, as it is the sum of the energies of one vortex
and one antivortex; on the other hand, when they approach each
other they should annihilate, giving \emph{zero energy}. Hence, at
least at some distance, the energy of the V-AV pair must decrease
as $d$ approaches zero and, as a result, the interaction potential
is attractive. This result is in contradiction with Ref.
\cite{Moshchalkov1}, where it was claimed that in type-I
superconductors the interaction of a V-AV pair is repulsive, and
which was used to explain the existence of a stable V-AV molecule
in mesoscopic superconducting triangles \cite{Moshchalkov2,
Teniers}. As follows from our theory, the V-AV interaction is
always attractive and should lead to a disfavored V-AV molecule in
type-I superconducting polygons, because in that case the V-V
interaction is also attractive. \cite{Golib, Roeland1}

One more question arises from the conjecture of Ref.
\cite{Moshchalkov1}: if vortices attract (repel) each other in
type-I (type-II) superconductors, whereas exactly the opposite
occurs for V-AV interactions, what one would expect in a
critically coupled system? In this regime, i.e., when $\mu = 1$,
vortices do not interact; should we expect the same for V-AV? The
answer to this question is provided by the Bogomol'nyi equations
\cite{Bogomolni}
\begin{subequations}
\begin{eqnarray}
&& \left[\left(\frac{\partial}{\partial x_1}-iA_1\right) + sgn(n)i\left(\frac{\partial}{\partial x_2}-iA_2\right)\right]\Psi = 0, \\
&& |\overrightarrow{\nabla}\times\overrightarrow{A}| + sgn(n)
\frac{1}{2}\left(|\Psi|^2-1\right) = 0.
\end{eqnarray}
\end{subequations}
For the single vortex ansatz chosen before, $\Psi(r, \theta) =
f(r)e^{i n \theta}$ and $\overrightarrow{A} =
na(r)\widehat{\theta}/r$, these equations read \cite{Samols}
\begin{subequations}\label{bogomolCIRC}
\begin{eqnarray}
&& r\frac{df}{dr} - sgn(n) n(1-a)f= 0 \\
&& \frac{2n}{r}\frac{da}{dr} + sgn(n)(f^2-1)= 0.
\end{eqnarray}
\end{subequations}
Substituting the latter formulae in Eq. (\ref{bogomolCIRC}) and
neglecting higher-order terms in $\sigma(r)$ and $Q(r)$ yields $nQ
= -sgn(n) r\frac{d\sigma}{dr}$ or, using the asymptotic forms of
these functions, $n\gamma_2 r K_1(r) = -sgn(n) \gamma_1 r
dK_0(r)/dr = sgn(n) \gamma_1 r K_1(r) \Rightarrow \gamma_1 =
sgn(n) n\gamma_2$. Substituting this expression for
$\gamma_1^{(i)}$ in Eq. (\ref{Eint}) shows that the interaction
potential for V-AV in the critical coupling regime is $E_{int}(d)
= 4n_1\gamma_2^{(1)}n_2\gamma_2^{(2)}K_0(d)$, which is still
attractive since $n_1n_2 < 0$. Hence, unlike vortex-vortex pairs,
a V-AV pair exhibits an attractive interaction even in the
critical case of $\mu = 1$.

It should be mentioned that, as $\gamma_1 \ne \gamma_2$ for $\mu
\ne 1$, the interaction potential given by Eq. (\ref{Eint}) may
diverge at small distances for some values of $\mu$. This signals
the breakdown of this analytical expression for small $d$.
Actually, for small separation $d$, the superposition ansatz
proposed by Bettencourt and Rivers and used in the present section
also fails, since it does not take into account either the spatial
deformation of the fields, or the possibility of the formation of
giant vortices. \cite{Jacobs} Hence, the analytical expression for
the interaction potential between vortices has a validity
restricted to large $d$. Nevertheless, when the interaction
potential is calculated by numerical means, taking into account
all the features mentioned above, the result shows very good
agreement with Eq. (\ref{Eint}) for $d$ larger than a critical
separation $d_c$ which depends on $\mu$, as will be shown in the
next section.

The interaction force $\Omega$ can be obtained by taking the
derivative of the energies with respect to the distance between
the two vortices. The results for $\mu = 1$ in the V-V and V-GV
cases clearly give $\Omega (d) = 0$. The analytical expression for
the force for large vortex-vortex separation can be easily derived
from Eq. (\ref{Eint}) as
\begin{equation}
\Omega (d) =
2\left[n_1n_2\gamma_2^{(1)}\gamma_2^{(2)}K_1(d)-\gamma_1^{(1)}\gamma_1^{(2)}\mu
K_1(\mu d)\right]. \label{Fint}
\end{equation}

\section{GL equations for fixed vortices} \label{numEq}

We recall step i) in the procedure described in the previous
section for obtaining the V-V interaction, which is fixing a
circular phase change around each vortex. In the present section,
we derive the Euler-Lagrange equations from Eqs. (\ref{EL1}) and
(\ref{EL2}) with the constraints imposed by step i).

In the paper by Jacobs and Rebbi \cite{Jacobs}, the authors fixed
the phase for a single vortex and obtained `modified' GL
equations, given by Eq. (2.18) of their paper or, equivalently,
Eqs. (\ref{ELcirc1}) and (\ref{ELcirc2}) of the present work, as
well as Eq. (7) of the paper by Babaev and Speight \cite{Babaev}.
Although they presented these equations in their paper, Jacobs and
Rebbi did not solve them directly, but used instead variational
functions for $f(r)$ and $a(r)$ and minimized the energy $E$
without solving the differential equations.

For the two vortex system, Jacobs and Rebbi made a different
ansatz, $\Psi = \exp[i \theta_1]\exp[i \theta_2]f(r,\theta)$,
where $\theta_1$ and $\theta_2$ are azimuthal angles around each
vortex position, $f$ is a real function which is not necessarily
circularly symmetric and is zero at the position of each vortex,
and the winding numbers were chosen as 1 for each vortex. Having
fixed the positions and vorticities, they just needed to find $f$
and $\vec{A}$ that minimize $E$. As before, instead of deriving
Euler-Lagrange (differential) equations, they used a variational
procedure, considering trial functions that account for the
deformation of the vortices towards the formation of the giant
vortex.

The results of Jacobs and Rebbi are rather accurate and the
advantage of their approach is that many terms of the variational
function can be integrated analytically. However, the variational
procedure involves many parameters, it is a very long analytical
calculation and their trial function is not the most general
function: if one wants to solve the problem for a V-AV or a V-GV
pair, the trial function has to be modified and consequently also
the analytical integrals in the variational procedure.

In order to obtain the inter-vortex potential, our approach starts
from the ansatz for two vortices $\Psi = e^{i n_1 \theta_1}e^{i
n_2 \theta_2}f(x,y)$, where we control the vorticity $n_1$ and
$n_2$ of each vortex. We further rewrite $e^{i n_j \theta_j}$ in
Cartesian coordinates:
\begin{equation}
e^{i n_j \theta_j} = \left(\frac{x_j + iy_j}{x_j -
iy_j}\right)^{n_j/2},
\end{equation}
where $\vec{r}_{j} = (x_j, y_j, 0) $ is the in-plane position
vector with origin in the center of the vortex $j$. As we will
study the case for two vortices separated by a distance $d$, we
take $\vec{r}_{1} = (x - d/2, y, 0) $ and $\vec{r}_{2} = (x + d/2,
y, 0)$.

Next, we substitute this ansatz into Eq. (\ref{lagrangeanUnits})
to get the energy functional $F$ for fixed position of vortices as
\begin{eqnarray}
F = \frac{1}{2} \left[\left(\frac{\partial f}{\partial
x}\right)^2 + \left(\frac{\partial f}{\partial y}\right)^2\right]\nonumber \hspace{2.5 cm} \\
+ \frac{1}{2}f^2 \left[ \overline{X}^2 + \overline{Y}^2 +
2(A_x\overline{Y}-A_y\overline{X}) + A^2\right] \nonumber \\ +
\frac{\mu^2}{8}(1-f^2)^2 + \frac{1}{2}|\vec{\nabla} \times
\vec{A}|^2, \hspace{2.1 cm} \label{lagrangean2}
\end{eqnarray}
where
\begin{equation}
\overline{X} = \frac{n_1 x_1}{r_1^2}+ \frac{n_2 x_2}{r_2^2},
\quad\quad\quad \overline{Y} = \frac{n_1 y_1}{r_1^2}+ \frac{n_2
y_2}{r_2^2}. \nonumber \label{eqx}
\end{equation}
Notice that although $\overline{X}$ and $\overline{Y}$ seem to
have no physical meaning, they can be related to the gauge
proposed by Jacobs and Rebbi for the vector potential, $\vec{A} =
na(r)\widehat{\theta}/r$, or $\vec{A} = (-n \sin\theta/r, n
\cos\theta/r, 0) = (-n y/r^2, n x/r^2, 0)$ in Cartesian
coordinates, which was shown to be compatible with the symmetry of
the Euler-Lagrange equations, leading to Eqs. (\ref{ELcirc1}) and
(\ref{ELcirc2}) for a single vortex. However, as we are looking
for general differential equations for two vortices, we will not
make any a priori choice of gauge for the vector potential.

In previous works, \cite{Jacobs, Babaev} the Euler-Lagrange
equations were not explicitly derived. We derived the
Euler-Lagrange equations for the present problem setting, which
are given by
\begin{subequations}
\begin{equation}
\nabla^2f - \left[\overline{X}^2 + \overline{Y}^2 +
2(A_x\overline{Y}-A_y\overline{X})+ A^2\right]f  +
\frac{\mu^2}{2}(1-f^2)f = 0, \label{EL3}
\end{equation}
and
\begin{equation}
\overrightarrow{\nabla} \times \overrightarrow{\nabla} \times
\overrightarrow{A} + \left[ \overrightarrow{A} - \frac{n_1
\widehat{\theta}_1}{r_1} - \frac{n_2 \widehat{\theta}_2}{r_2}
\right]f^2 = 0, \label{EL4}
\end{equation}
\end{subequations}
where the unitary angular vectors around each vortex can be
rewritten as $\widehat{\theta_j} = (-y_j/r_j, x_j/r_j, 0)$. One
can even verify, after some manipulations of the equations, that
inserting $n_2 = 0$ in these equations leads to Eq. (2.18) in the
paper by Jacobs and Rebbi, which is the equation for a single
vortex, but in Cartesian coordinates. Solving Eqs. (\ref{EL3}) and
(\ref{EL4}) is equally demanding as solving the common GL
equations, which was done in many works in the literature.
\cite{Schweigert, Golib, Roeland1}

\section{Numerical results and fitting functions}

We solved Eqs. (\ref{EL3}) and (\ref{EL4}) numerically using the
finite difference technique and a relaxation method suitable for
non-linear differential equations. \cite{NR} The two-dimensional
system is divided in a uniform square 601$\times$601 grid with
total dimensions $60\lambda \times 60\lambda$. The singularities
in the amplitude of the order parameter appear naturally in the
center of each vortex position, as a consequence of the fixed
circular phase $e^{i n_i \theta_i}$ defined around each vortex
$i$, which guarantees the existence of zeros of the order
parameter in the center of the vortices. \cite{Jacobs} The results
for V-V, V-GV and V-AV interactions are presented separately in
the following subsections. Analytical fitting functions will be
proposed for the numerically obtained curves, where the fitting
error is defined by the variance \cite{NR}
\begin{equation}
\nu =
\sum_{n=1}^{N}\frac{\left[G(n)-G_{fit}(n)\right]^2}{(N-N_p)},
\end{equation}
where $G(n)$ is the numerical data set, $G_{fit}(n)$ is the
analytical fitting function, $N$ is the length of the data set and
$N_p$ is the number of variational parameters of the fitting
function. \cite{foot}

\subsection{Vortex-vortex interaction}

\begin{figure}[!t]
\centerline{\includegraphics[width=\linewidth]{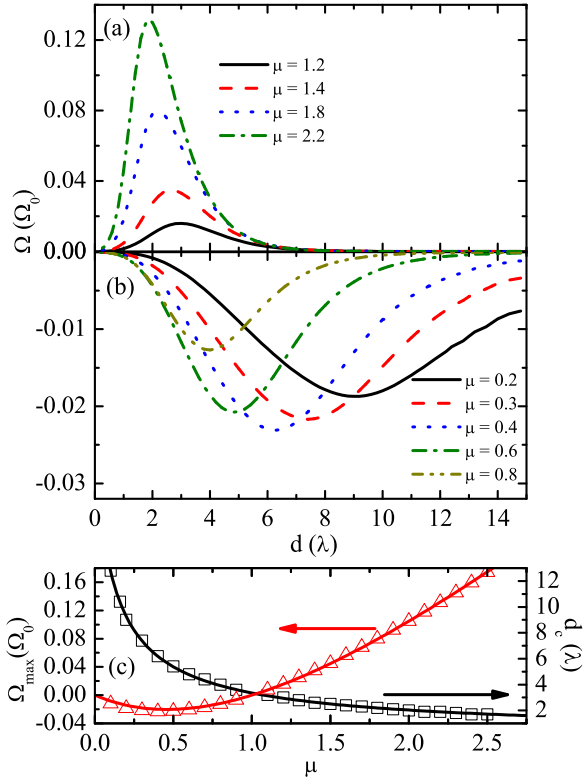}}
\caption{(Color online) Numerically obtained vortex-vortex
interaction force $\Omega$ as a function of the separation $d$
between vortices, for several values of $\mu = \sqrt{2}\kappa$ in
the (a) type-II and (b) type-I regimes. (c) Critical separation
$d_c$ (open squares, right scale) and extremum $\Omega_{max}$
(open triangles, left scale), which correspond respectively to the
position and amplitude of the peak in the force, as a function of
the GL parameter. The fitting functions for $d_c$ and
$\Omega_{max}$ are plotted by the solid curves.} \label{fig:1}
\end{figure}

The numerical results for the V-V interaction force are shown in
Fig. \ref{fig:1}, for several values of $\mu$ in the type-II (a)
and type-I (b) regimes. Notice that for $\mu \approx 0$ vortices
should not interact and the force vanishes, but for $0< \mu < 1$
they attract (type-I regime) and the force is negative in this
case. However, at the critical point $\mu = 1$ the force vanishes
again. Hence, in the type-I case, two different regimes can be
identified: one where the force increases from zero, as $\mu$
increases from zero, and the other when the force decreases back
to zero, as $\mu$ approaches 1. This can be seen from Fig.
\ref{fig:1}(c), where extremum of the force peak $\Omega_{max}$
(open triangles, left scale) increases with $\mu$ for small $\mu$,
but for $0.6 < \mu < 1$ the peak decreases with $\mu$, approaching
zero when $\mu = 1$. The numerical results can be fitted to
$\Omega_{max} (\mu) = 0.0961\mu(\mu-1)\big/(1+0.2863\mu)^{1.341}$,
which is shown by the red curve in Fig. \ref{fig:1}(c). From Figs.
\ref{fig:1}(a,b) we see that the force exhibits a maximum at some
critical separation $d_c$, which depends on the GL parameter $\mu
= \sqrt{2}\kappa$. The critical separation $d_c$ is also shown in
Fig. \ref{fig:1}(c) (open squares, right scale) as a function of
$\mu$. The fit of this curve, given by $d_c =
22.203(1+10.504\mu)^{-0.774}$ (with estimated variance $\nu
\approx 0.3\%$), is shown by the solid curve in Fig.
\ref{fig:1}(c), which suggests that the critical separation for
the V-V interaction approaches zero in the extreme type-II
situation ($\mu \rightarrow \infty$). An effective extreme type-II
scenario can also be achieved in a superconducting film of
thickness $w \ll \lambda$, where the effective penetration depth
is $\Lambda = \lambda^2 / w$, and for which analytical expressions
for the V-V interaction force were proposed by Pearl \cite{Pearl},
and later by Brandt \cite{Brandt3}. However, in the case of
superconducting films, the V-V force decays monotonically as
$1/d^2$, while in the present case of a bulk superconductor the
decay is exponential. Hence, although both situations can be
considered as extreme type-II limits, our results for bulk
superconductors with $\mu \rightarrow \infty$ are quantitatively
different from those for thin superconducting films.

\begin{figure}[!b]
\centerline{\includegraphics[width=\linewidth]{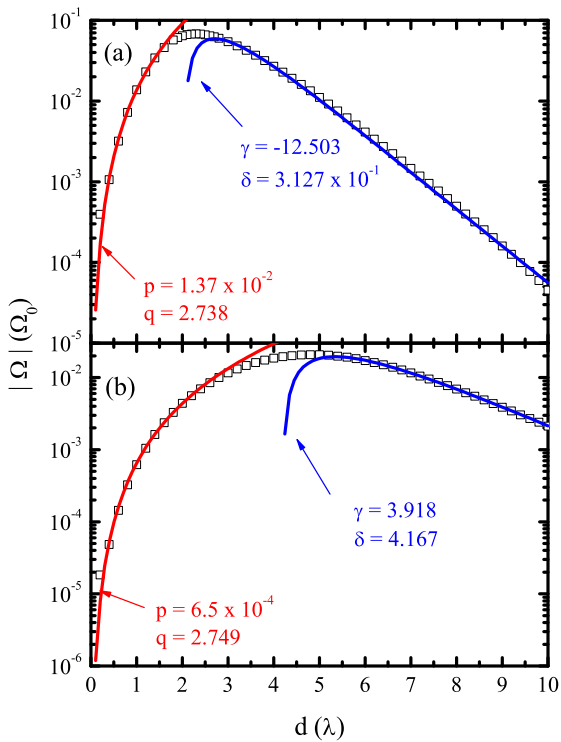}}
\caption{(Color online) Comparison between the V-V interaction
force as a function of the separation $d$ obtained by the
numerical method (symbols) and by the asymptotic expressions
(curves) from Eqs. (\ref{limfinal}, \ref{limbeg}), for $\mu = 1.7$
(a) and $\mu = 0.6$ (b). The forces are plotted on a $log_{10}$
scale and the values of the fitting parameters $p, q, \gamma$ and
$\delta$ are given in each panel.}\label{fig:3}
\end{figure}

The importance of solving Eqs. (\ref{EL3}) and (\ref{EL4})
numerically for two separate vortices lies in the possibility of
obtaining the interaction force between vortices even in the small
separation limit, which cannot be described by the asymptotic
functions given in the literature \cite{Bettencourt} and described
by Eq. (\ref{Fint}). However, since solving these equations is
generally not an easy task, we attempt to propose here an
analytical expression that possesses all the features of the
numerically obtained force as a function of the vortex separation.
Such an analytical expression can be helpful e.g. for numerical
modelling of vortex structures by means of molecular dynamics (MD)
simulations, where vortices are considered as point particles. Up
to now, in those MD simulations one typically uses the asymptotic
analytical expressions for the V-V interaction which are only
valid in the large separation limit. \cite{Auto, Brandt2} To
obtain a proper fitting function, we first analyze separately the
behaviors for large and small vortex separations. For large
separation, using the asymptotic form of the modified Bessel
functions, Eq. (\ref{Fint}) can be rewritten as
\begin{equation}
\Omega (d \rightarrow \infty) = \gamma
d^{-\frac{1}{2}}\left(\delta e^{-d}-\sqrt{\mu}e^{-\mu d}\right),
\label{limfinal}
\end{equation}
where $\gamma$ and $\delta$ are fitting parameters. For small
separation, our results show that a power function of $d$
describes the force satisfactorily, i.e.
\begin{equation}
\Omega (d \rightarrow 0) =  pd^{q}, \label{limbeg}
\end{equation}
with $p$ and $q$ as fitting parameters. Two examples of such
fittings are shown in Fig. \ref{fig:3}, for $\mu$ = 0.6 (bottom)
and $\mu =$ 1.7 (top). Notice that the parameter $q$ depends
weakly on $\mu$, exhibiting values between $\approx 2.7$ and
$\approx 2.8$ for all values of $\mu$ considered in the V-V case.
\begin{figure}[!t]
\centerline{\includegraphics[width=\linewidth]{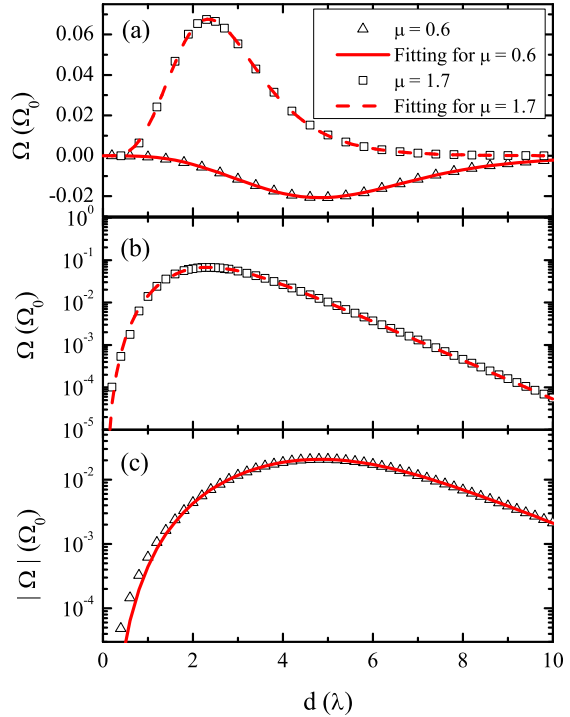}}
\caption{(Color online) (a) Comparison between the V-V interaction
force as a function of the separation $d$ obtained by the
numerical method (symbols) and by the fitting function (curves)
given by Eq. (\ref{compfit}). The solid (dashed) curves and the
open triangles (squares) are the results for $\mu = 0.6$ (1.7).
(b-c) The results for each $\mu$ are plotted separately in a
log-scale, to emphasize the maximal deviation of the fitting
function from the data.}\label{fig:4}
\end{figure}

Following the established behavior of the interaction in the
limiting cases, we propose a single function which has the above
limits as limiting behaviors:
\begin{equation}
\Omega_{fit}(d) = \eta_1 \frac{d^{\eta_3}}{1+\eta_2
d^{\eta_3+\frac{1}{2}}}\left(\eta_4 e^{-d}-\sqrt{\mu}e^{-\mu
d}\right), \label{compfit}
\end{equation}
where $\eta_i$ ($i$ = 1 - 4) are four fitting parameters.

\begin{figure}[!b]
\centerline{\includegraphics[width=\linewidth]{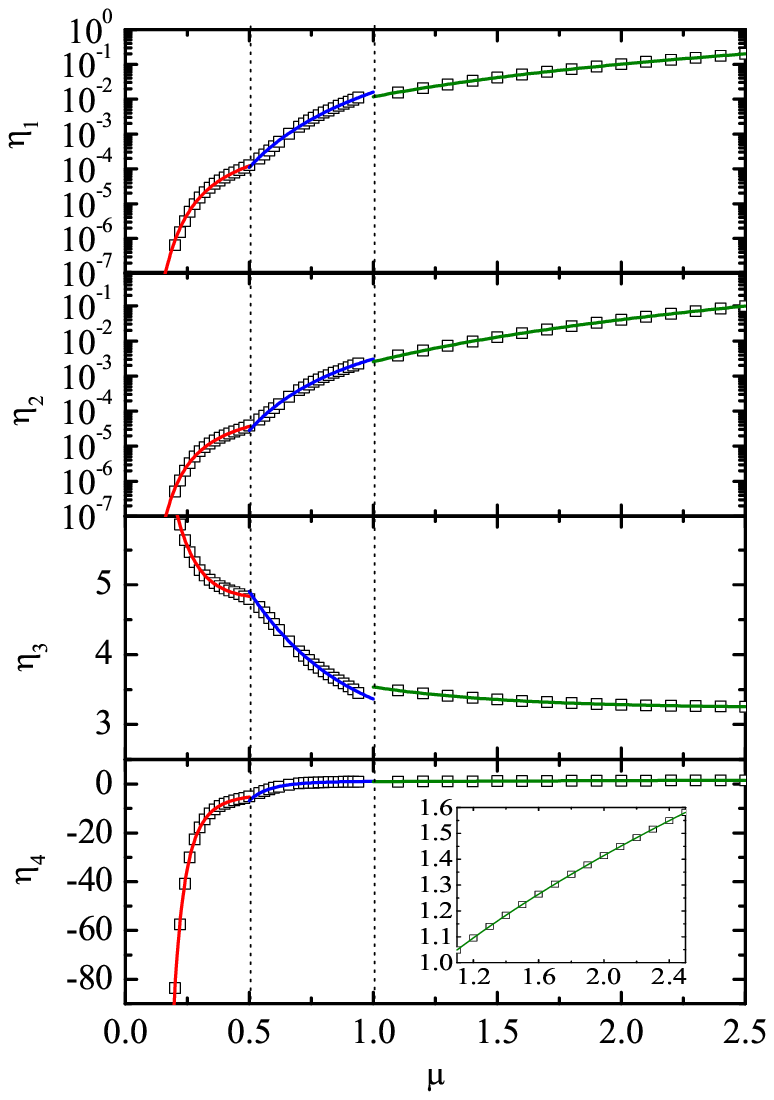}}
\caption{(Color online) Fitting parameters (symbols) in Eq.
(\ref{compfit}) as a function of $\mu = \sqrt{2}\kappa$ for the
V-V case. The curves are the fitting functions given by Eqs.
(\ref{eqPar}) for $\eta_i (\mu)$ ($i = 1 - 4$), in three different
regions: $0 < \mu < 0.5$,  $0.5 < \mu < 1$ and $1 < \mu$. The
inset shows a magnification of the results for $\eta_4$ at large
$\mu$.}\label{fig:5}
\end{figure}

Fig. \ref{fig:4} shows the fitting obtained with Eq.
(\ref{compfit}) for the same values of $\mu$ presented in Fig.
\ref{fig:3}. The fitting is not ideal for $d < \lambda$, where the
force becomes very small. Nevertheless, we found that the fitting
error is lower than $1 \%$. Please note that the V-V interaction
potential, which is an integrated force, will be even more
accurate.

The values of the four fitting parameters are given in Table I,
for $\mu$ from 0.2 to 2.5. Notice that the estimated variance
$\nu$ increases with $\mu$, and thus Eq. (\ref{compfit}) should
not be used in the extreme type-II case. Nevertheless, in the
extreme type-II case the critical separation $d_c$ approaches
zero, as mentioned earlier, and consequently, the short range part
of the V-V interaction force will not be important in such a
situation. Hence, the asymptotic expression $\Omega(d) = f_0
K_1\left(d\right)$ frequently used in the literature, \cite{Nori,
Pogosov} which can be obtained from Eq. (\ref{Fint}) by making
$\mu \rightarrow \infty$, is expected to provide a good
description of the V-V interaction force in extreme type-II
situations.

\begin{table*}
\caption{Fitting parameters $\eta_i$ and the estimated variance
$\nu$ for Eq. (\ref{compfit}) in the V-V case, for different
values of $\mu$.} \label{tab:1}\renewcommand{\arraystretch}{1}
\begin{ruledtabular}
\begin{tabular}{cccccc}
  \hline
  $\mu$ & $\eta_1$ & $\eta_2$ & $\eta_3$ & $\eta_4$ & $\nu (\times 10^{-8})$ \\
  \hline
 0.2 & 6.564$\times10^{-7}$ & 5.13$\times10^{-7}$ & 6.135 & -83.611 & 1.42\\
 0.3 & 1.522$\times10^{-5}$ & 7.268$\times10^{-6}$ & 5.213 & -17.667 & 3.96 \\
 0.4 & 5.698$\times10^{-5}$ & 2.041$\times10^{-5}$ & 4.950 & -8.014 & 5.06 \\
 0.5 & 1.284$\times10^{-4}$ & 3.911$\times10^{-5}$ & 4.796 & -5.090 & 1.95 \\
 0.6 & 4.474$\times10^{-4}$ & 1.222$\times10^{-4}$ & 4.440 & -1.538 & 0.448 \\
 0.7 & 1.62$\times10^{-3}$ & 3.968$\times10^{-4}$ & 4.046 & 0.237 & 0.239 \\
 0.8 & 4.12$\times10^{-3}$ & 9.326$\times10^{-4}$ &  3.760 & 0.766 & 0.215 \\
 0.9 & 8.46$\times10^{-3}$ & 1.79$\times10^{-3}$ & 3.544 & 0.943 &0.116\\
 1.1 & 1.546$\times10^{-2}$ & 3.88$\times10^{-3}$ & 3.489 & 1.049 &0.139\\
 1.2 & 2.068$\times10^{-2}$ & 5.37$\times10^{-3}$ & 3.443 & 1.095 &0.98\\
 1.3 & 2.67$\times10^{-2}$ & 7.31$\times10^{-3}$ & 3.410 & 1.140 &3.16\\
 1.4 & 3.369$\times10^{-2}$ & 9.78$\times10^{-3}$ & 3.382 & 1.183 &7.24\\
 1.5 & 4.175$\times10^{-2}$ & 1.286$\times10^{-2}$ & 3.358 & 1.225 &13.8\\
 1.6 & 5.094$\times10^{-2}$ & 1.664$\times10^{-2}$ & 3.338 & 1.265 &23.6\\
 1.7 & 6.136$\times10^{-2}$ & 2.121$\times10^{-2}$ & 3.320 & 1.304 &36.8\\
 1.8 & 7.308$\times10^{-2}$ & 2.667$\times10^{-2}$ & 3.306 & 1.342 &53.9\\
 1.9 & 8.618$\times10^{-2}$ & 3.311$\times10^{-2}$ & 3.294 & 1.378 &75.3\\
 2.0 & 0.1008 & 4.066$\times10^{-2}$ & 3.283 & 1.414 &101.1\\
 2.1 & 0.1169 & 4.94$\times10^{-2}$ & 3.275 & 1.449 &131.6\\
 2.2 & 0.1347 & 5.945$\times10^{-2}$ & 3.268 & 1.483 &166.7\\
 2.3 & 0.1542 & 7.093$\times10^{-2}$ & 3.262 & 1.517 &206.6\\
 2.4 & 0.1756 & 8.396$\times10^{-2}$ & 3.258 & 1.549 &251.1\\
 2.5 & 0.199 & 9.866$\times10^{-2}$ & 3.254 & 1.581 &286.2\\
  \hline
\end{tabular}
\end{ruledtabular}
\end{table*}

We next attempt to find an analytical expression for the fitting
parameters as function of $\mu$. Their dependence on $\mu$ is
shown in Fig. \ref{fig:5}. Three different ranges of $\mu$,
delimited by vertical dotted lines in Fig. \ref{fig:5}, can be
distinguished. The physical reason for the existence of three
different behaviors of the parameters $\eta_i$ as a function of
$\mu$ is the following: for type-I ($\mu < 1$), as we explained
earlier, there must be a regime where the size of the attractive
force peak increases with $\mu$ and another region where it
decreases with $\mu$. This defines the ranges 1 ($\mu < 0.5$) and
2 ($0.5 < \mu < 1$), respectively. Range 3 is then the type-II
regime, for $\mu
> 1$, where the interaction force is repulsive. The functions
$\eta_i (\mu)$ in Fig. \ref{fig:5} were fitted as
\begin{subequations}\label{eqPar}
\begin{equation}
\eta_1 (\mu) = e^{B_1(\mu^{C_1}+A_1)},
\end{equation}
\begin{equation}
\eta_2 (\mu) = e^{B_2(\mu^{C_2}+A_2)},
\end{equation}
\begin{equation}
\eta_3(\mu) = A_3 + B_3e^{C_3\mu}
\end{equation}
and
\begin{equation}
\eta_4(\mu) = A_4 + B_4\mu^{C_4},
\end{equation}
\end{subequations}
with different parameters $A_i, B_i$ and $C_i$ for each range
listed in Table II. These fitting functions for $\eta_i (\mu)$ are
shown as solid curves in Fig. \ref{fig:5}. Notice that the
parameter $\eta_4$ must satisfy the condition $\eta_4 \le
\sqrt{\mu}$ ($\ge \sqrt{\mu}$) in the type-I (type-II) case,
otherwise the difference between the exponential terms in Eq.
(\ref{compfit}) would exhibit a sign change for small separations,
leading to a spurious repulsive (attractive) region in this case.
In the type-II case, this condition leads to $\eta_4 (\mu) \approx
\sqrt{\mu}$ as the best value for this fitting parameter.

\begin{table}
\caption{Fitting parameters in Eqs. (\ref{eqPar} a-d) for the V-V
case, for three different ranges of $\mu$.}
\label{tab:2}\renewcommand{\arraystretch}{1}
\begin{ruledtabular}
\begin{tabular}{cccc}
  \hline
 Parameter & $\mu < 0.5$ & $0.5 < \mu < 1$ & $\mu > 1$ \\
  \hline
 $A_1$  & 5.977 & -0.5420 & -0.9404\\
 $B_1$  & -1.092 & -9.041 & -74.584 \\
 $C_1$  & -1.191 & -0.6323 & -4.221$\times10^{-2}$ \\
 $A_2$  & 13.845 & 7.935 $\times10^{-2}$ & -0.9843\\
 $B_2$  & -0.6218 & -5.359 & -379.321 \\
 $C_2$  & -1.373 & -0.9084 & -1.057$\times10^{-2}$ \\
 $A_3$  & 4.79 & 2.756 & 3.234\\
 $B_3$  & 12.542 & 7.587 & 1.849 \\
 $C_3$  & -11.183 & -2.523 & -1.804 \\
 $A_4$  & -3.677 & 1.215 & 0\\
 $B_4$  & -8.663$\times10^{-2}$ & -0.1229 & 1 \\
 $C_4$  & -4.244 & -6.022 & 0.5 \\
  \hline
\end{tabular}
\end{ruledtabular}
\end{table}

It is important to point out that the results obtained for
$\eta_2$ are not the same as the values of $q$ in Eq.
(\ref{limbeg}) for the power law at small separations, which, as
mentioned earlier, are between $\approx 2.7$ and $\approx 2.8$.
This is reasonable, because the exponential terms in Eq.
(\ref{compfit}) still play a role in the small $d$ limit of this
expression, thus, the parameter $\eta_2$ must assume a value that
is different from $q$ in order to compensate these terms. The
values obtained for $\eta_4$, which is the parameter controlling
the large $d$ range of Eq. (\ref{Fint}), are also not the same as
the values obtained when one uses the asymptotics of each single
vortex to find the parameters $\gamma_1^{(i)}$ and
$\gamma_2^{(i)}$ in Eq. (\ref{Fint}). Actually, for $\mu > 1$, we
found $\eta_4 \approx \sqrt{\mu}$, which is equivalent as making
$\gamma_1^{(i)} = \gamma_2^{(i)}$. As $\mu$ increases, the
difference between $\gamma_1^{(i)}$ and $\gamma_2^{(i)}$ is
amplified \cite{Speight}, leading to a higher variance $\nu$ for
large $\mu$, as shown in Table I. Even so, this choice of $\eta_4$
conveniently leads to a function which decays exponentially for
large separation $d$, as expected for V-V interactions in bulk
superconductors, and which exhibits no sign change at small
separations. Of course, the fitting function Eq. (\ref{compfit})
can be improved to provide a better fitting of the large
separation part and to reproduce a perfect power law for small
separations, but this would require more fitting parameters and
very complicated expressions. Equation (\ref{compfit}) is simple
and still accurate for $0 \le \mu \le 2.5$, as verified by the
small variances $\nu < 10^{-6}$ in Table I and by the comparison
with the numerical results in Fig. \ref{fig:4}.

\subsection{Vortex-giant vortex interaction}

\begin{figure}[!t]
\centerline{\includegraphics[width=\linewidth]{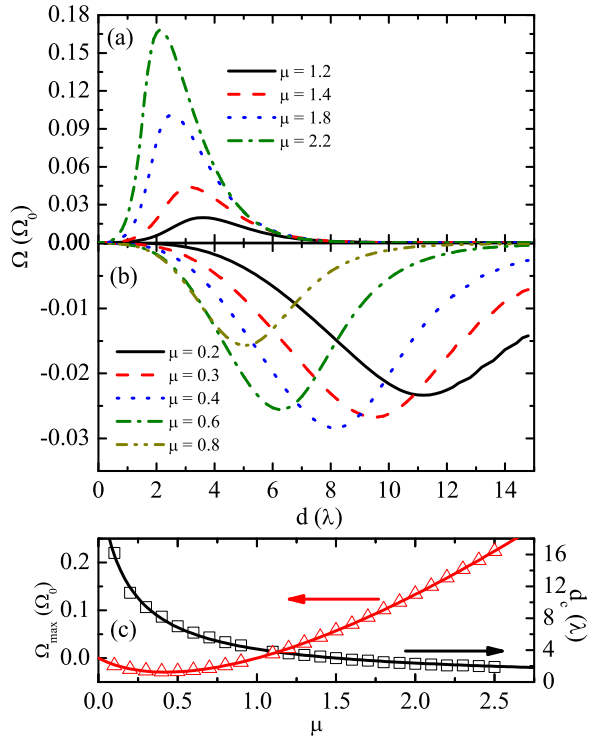}}
\caption{(Color online) Numerically obtained V-GV interaction
force $\Omega$ as a function of the separation $d$ between a
vortex and a double-quantized vortex, for several values of $\mu =
\sqrt{2}\kappa$ in the (a) type-II and (b) type-I regimes. (c)
Critical separation $d_c$ (open squares, right scale) and extremum
$\Omega_{max}$ (open triangles, left scale), which correspond
respectively to the position and amplitude of the peak in the
force, as a function of the GL parameter. The fitting functions
for $d_c$ and $\Omega_{max}$ are plotted by the solid
curves.}\label{fig:6}
\end{figure}

As observed in Fig. \ref{fig:1}, the interaction force between two
vortices shows a maximum at some critical separation $d_c$ and
decays to zero for both very large and very small separations. The
former is reasonable, since the interaction between vortices is
expected to weaken as they are placed further from each other. The
latter is due to the formation of a giant vortex state: when two
vortices of winding numbers e.g. $n_1 = 1$ and $n_2 = 1$ are put
close to each other, they coalesce, forming a $n = n_1 + n_2 = 2$
giant vortex. \cite{Jacobs, Schweigert, Kanda, Golubovic} In the
absence of lateral confinement, a giant vortex is a stable
(unstable) state in type-I (type-II) systems and can interact as
such with other vortices, and this motivated us to investigate the
interaction force between a vortex and a giant vortex.

The V-GV interaction force is shown in Fig. \ref{fig:6} as a
function of the distance between them, for several values of $\mu$
in the type-II (a) and type-I (b) regimes. The behavior of the
curves is quite similar to those in Fig. \ref{fig:1} for the V-V
case, but with different amplitudes and critical separations. The
critical separation $d_c$, beyond which the vortices start to
coalesce, obtained numerically for the V-GV interaction is shown
as a function of the GL parameter $\mu$ in Fig. \ref{fig:6} (c),
along with its fitting function $d_c = 25.043(1+6.632
\mu)^{-0.8862}$ (with estimated variance $\nu \approx 2\%$).
Notice that the critical separation for the V-GV interaction is
always larger than the one for the V-V case, because the giant
vortex has a larger core in comparison to a $n = 1$ vortex.
Nevertheless, the fitting function shows that the smallest
critical separation for the V-GV interaction force, which would be
obtained in an extreme type-II regime, is also $d_c (\mu
\rightarrow \infty) = 0$, as in the V-V case. The behavior of the
extremum of the force peak $\Omega_{max}$ as a function of $\mu$,
shown as open triangles in Fig. \ref{fig:6}(c), is similar to that
found for the V-V case, with the amplitude approaching zero for
$\mu \rightarrow 0$ and $\mu \rightarrow 1$, and increasing
monotonically for $\mu$ increasing above 1. The extremum of the
force peak can be fitted to $\Omega_{max} =
0.1709\mu(\mu-1)/(1+1.854\mu)^{0.6087}$, which is shown by the
solid curve in Fig. \ref{fig:6}(c).

\begin{figure}[!t]
\centerline{\includegraphics[width=\linewidth]{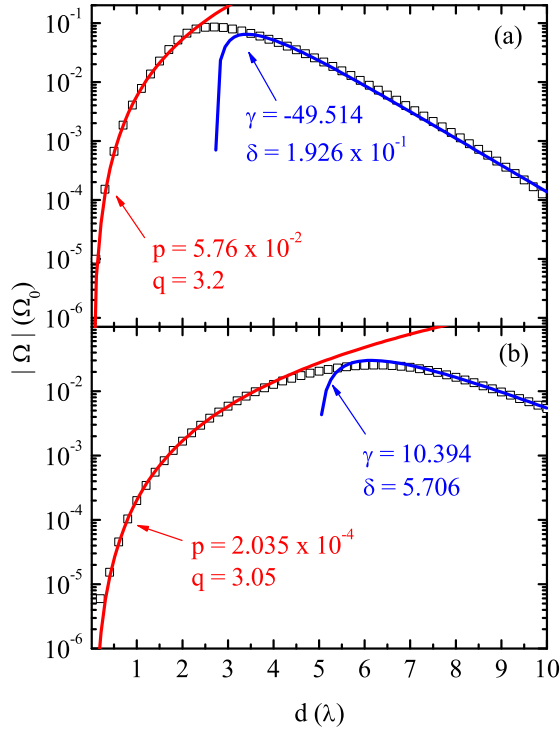}}
\caption{(Color online) Comparison between the V-GV interaction
force as a function of the separation $d$, obtained by the
numerical method (symbols) and by the asymptotic expressions
(curves) from Eqs. (\ref{limfinal}, \ref{limbeg}), for $\mu = 1.7$
(a) and $\mu = 0.6$ (b). The forces are plotted on a $log_{10}$
scale and the values of the fitting parameters $p, q, \gamma$ and
$\delta$ are given in each panel.}\label{fig:8}
\end{figure}

In Sec. II, we analytically found that Eq. (\ref{Fint}) remains
valid for the asymptotic V-GV interactions, simply by choosing
$n_1 = 1$, $n_2 = 2$ and changing the parameters $\gamma_1^{(i)}$
and $\gamma_2^{(i)}$ accordingly. Moreover, our results show that
the force in the small separation limit in this case can still be
well described by a power function of the separation $d$. The
fitting of the force for the small and large separation limits,
given by Eqs. (\ref{limfinal}, \ref{limbeg}), are shown in Fig.
\ref{fig:8} for the V-GV interaction, agreeing well with the
numerical results. This suggests that the fitting function given
by Eq. (\ref{compfit}) can not only be used for the V-V, but also
for the V-GV interaction force.

\begin{figure}[!b]
\centerline{\includegraphics[width=\linewidth]{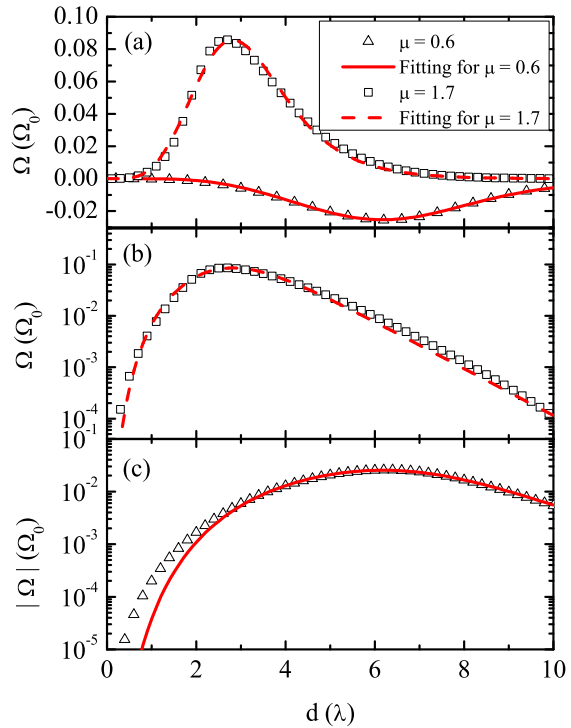}}
\caption{(Color online) (a) Comparison between the V-GV
interaction force as a function of the separation $d$ obtained by
the numerical method (symbols) and by the fitting function
(curves) given by Eq. (\ref{compfit}). The solid (dashed) curves
and the open triangles (squares) are the data for $\mu = 0.6$
(1.7). The results for each $\mu$ are also plotted separately on a
log-scale: (b) $\mu = 1.7$; (c) $\mu = 0.6$.}\label{fig:9}
\end{figure}

The four fitting parameters $\eta_i$ ($i$ = 1 - 4) found for each
value of $\mu$ in the V-GV case are shown in Table III, for $\mu$
from 0.2 to 2.5. As in the V-V case, the estimated variance $\nu$
increases for $\mu > 1$, hence the offered function is expected to
fail in the extreme type-II case. As an example, the V-GV
interaction force for $\mu = 1.7$ and 0.6 is shown in Fig.
\ref{fig:9} as a function of the vortex-giant vortex separation
$d$, along with the fitting curves given by Eq. (\ref{compfit})
with the corresponding parameters in Table III. Although the
estimated variances for these cases are smaller than $10^{-5}$, it
can be seen in the log-plot in Fig. \ref{fig:9}(b) and (c) that
for small separation the fitting function is less accurate as
compared to the V-V case shown in Fig. \ref{fig:4}, where the
variances are lower than $10^{-7}$. Nevertheless, in the low $d$
region the force is small and consequently the deviation in the
force will also be small.

\begin{table*}
\caption{Fitting parameters $\eta_i$ and estimated variance $\nu$
for Eq. (\ref{compfit}) in the V-GV case ($n_1 = 1$ and $n_2 =
2$), for $\mu$ between 0.2 and 2.5.}
\label{tab:3}\renewcommand{\arraystretch}{1}
\begin{ruledtabular}
\begin{tabular}{cccccc}
  \hline
  $\mu$ & $\eta_1$ & $\eta_2$ & $\eta_3$ & $\eta_4$ & $\nu (\times 10^{-8})$ \\
  \hline
 0.2 & 1.495$\times10^{-9}$ & 6.073$\times10^{-10}$ & 8.440 & -443.65 & 1.72\\
 0.3 & 2.347$\times10^{-7}$ & 4.944$\times10^{-8}$ & 6.929 & -58.765 & 16.2 \\
 0.4 & 1.165$\times10^{-6}$ & 1.69$\times10^{-7}$ & 6.679 & -24.312 & 33.7 \\
 0.5 & 2.302$\times10^{-6}$ & 2.823$\times10^{-7}$ & 6.628 & -19.433 & 22.6 \\
 0.6 & 1.083$\times10^{-5}$ & 1.230$\times10^{-6}$ & 6.199 & -8.372 & 8.6 \\
 0.7 & 6.775$\times10^{-5}$ & 6.892$\times10^{-6}$ & 5.681 & -1.661 & 2.6 \\
 0.8 & 3.681$\times10^{-4}$ & 3.283$\times10^{-5}$ &  5.173 & 0.446 & 0.91 \\
 0.9 & 1.54$\times10^{-3}$ & 1.196$\times10^{-4}$ & 4.706 & 0.942 & 0.41\\
 1.1 & 3.81$\times10^{-3}$ & 3.847$\times10^{-4}$ & 4.539 & 1.049 &0.69\\
 1.2 & 5.98$\times10^{-3}$ & 6.366$\times10^{-4}$ & 4.423 & 1.095 &6.0\\
 1.3 & 8.78$\times10^{-3}$ & 1.01$\times10^{-3}$ & 4.330 & 1.140 &21.7\\
 1.4 & 1.237$\times10^{-2}$ & 1.55$\times10^{-3}$ & 4.248 & 1.183 &54.0\\
 1.5 & 1.684$\times10^{-2}$ & 2.3$\times10^{-3}$ & 4.175 & 1.225 &108.8\\
 1.6 & 2.177$\times10^{-2}$ & 3.27$\times10^{-3}$ & 4.148 & 1.265 &196.0\\
 1.7 & 2.813$\times10^{-2}$ & 4.57$\times10^{-3}$ & 4.095 & 1.304 &315.8\\
 1.8 & 3.561$\times10^{-2}$ & 6.22$\times10^{-3}$ & 4.048 & 1.342 &474.8\\
 1.9 & 4.428$\times10^{-2}$ & 8.28$\times10^{-3}$ & 4.008 & 1.378 &676.0\\
 2.0 & 5.421$\times10^{-2}$ & 1.08$\times10^{-2}$ & 3.973 & 1.414 &676.0\\
 2.1 & 6.549$\times10^{-2}$ & 1.384$\times10^{-2}$ & 3.942 & 1.449 &1218\\
 2.2 & 7.818$\times10^{-2}$ & 1.747$\times10^{-2}$ & 3.916 & 1.483 &1564\\
 2.3 & 9.236$\times10^{-2}$ & 2.174$\times10^{-2}$ & 3.894 & 1.517 &1962\\
 2.4 & 0.1081 & 2.671$\times10^{-2}$ & 3.876 & 1.549 &2410\\
 2.5 & 0.1255 & 3.245$\times10^{-2}$ & 3.861 & 1.581 &2911\\
  \hline
\end{tabular}
\end{ruledtabular}
\end{table*}

\begin{figure}[!b]
\centerline{\includegraphics[width=\linewidth]{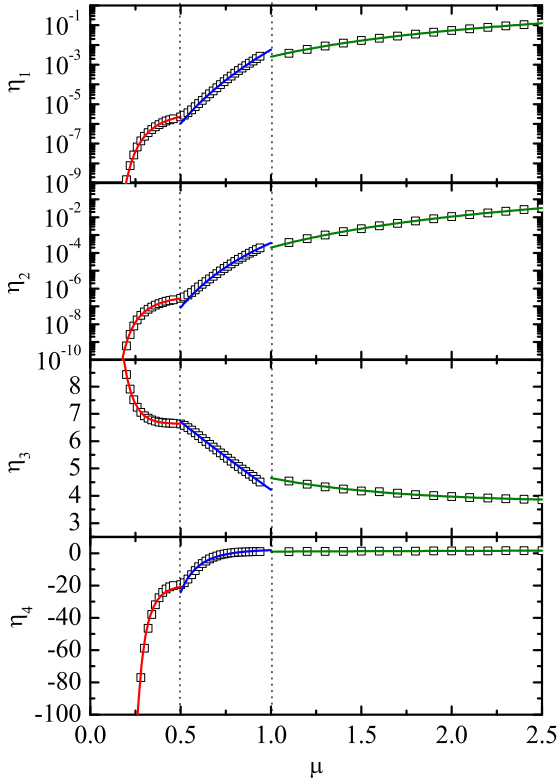}}
\caption{(Color online) The fitting coefficients $\eta_{i =
1-4}$(symbols) in Eq. (\ref{compfit}) as a function of $\mu =
\sqrt{2}\kappa$ for the V-GV case. The curves present the $\eta_i
(\mu)$ fittings given by Eqs. (\ref{eqPar}a-d) for three different
ranges, defined in the text as range 1 ($0 < \mu < 0.5$), range 2
($0.5 < \mu < 1$) and range 3 ($1 < \mu$).}\label{fig:12}
\end{figure}

The dependence of the fitting parameters on $\mu$ is shown in Fig.
\ref{fig:12} (squares) for the case of a $n_1 = 1$ and $n_2 = 2$
V-GV interaction, where the fitting to the data is also shown
(curves). Once more, the three different behaviors of $\eta_i$ as
a function of $\mu$ are observed, with data fitted with different
parameters $A_i, B_i$ and $C_i$, using Eqs. (\ref{eqPar}a-d). The
reason for the existence of three identifiable behaviors of the
parameters $\eta_i$ as a function of $\mu$ is the same as in the
V-V case. Moreover, for $\mu > 1$ (type-II regime), we found
$\eta_4 (\mu) \approx \sqrt{\mu}$, similar to the V-V case. The
parameters $A_i, B_i$ and $C_i$ for the V-GV case, for each range
of $\mu$, are given in Table IV.

\begin{table} [!b]
\caption{Fitting parameters in Eqs. (\ref{eqPar}a-d) for the V-GV
case ($n_1 = 1$ and $n_2 = 2$), for three different ranges of
$\mu$.} \label{tab:4}\renewcommand{\arraystretch}{1}
\begin{ruledtabular}
\begin{tabular}{cccc}
  \hline
 Parameter & $\mu < 0.5$ & $0.5 < \mu < 1$ & $\mu > 1$ \\
  \hline
 $A_1$  & 61.948 & -0.9926 & -0.5785\\
 $B_1$  & -0.1937 & -693.484 & -14.199 \\
 $C_1$  & -2.350 & -1.779$\times10^{-2}$ & -0.3514 \\
 $A_2$  & 101.28 & -0.6458 & -0.5461\\
 $B_2$  & -0.1417 & -22.365 & -18.823 \\
 $C_2$  & -2.374 & -0.4566 & -0.3461 \\
 $A_3$  & 6.631 & -15.424 & 3.761\\
 $B_3$  & 67.184 & 24.938 & 3.748 \\
 $C_3$  & -18.061 & -0.2385 & -1.442 \\
 $A_4$  & -19.425 & 2.654 & 0\\
 $B_4$  & -2.396$\times10^{-2}$ & -0.7764 & 1 \\
 $C_4$  & -6.075 & -5.113 & 0.5 \\
  \hline
\end{tabular}
\end{ruledtabular}
\end{table}

\subsection{The three vortex problem}

Up to now, we considered only the two-body interaction of
vortices. In the study of the dynamics of many vortices, one
generally considers the sum of pairwise interactions. In this
sense, the force acting on vortex $i$ in a system with many
vortices forming a certain configuration is given by \cite{Nori,
Pogosov}
\begin{equation}
\vec{\Omega}_i = \sum_{j \neq i}\Omega\left(|\vec{r_i} -
\vec{r_j}|\right) \hat{r}_{i,j}
\end{equation}
where $\vec{r_i}$ is the position of the vortex $i$ and
$\hat{r}_{i,j} = \left(\vec{r_i} - \vec{r_j}\right)/|\vec{r_i} -
\vec{r_j}|$. In such a model, the V-V interaction force is derived
from the interaction potential between a pair of vortices, and is
usually taken as $\Omega(|\vec{r_i} - \vec{r_j}|) = f_0
K_1\left(|\vec{r_i} - \vec{r_j}|/\lambda\right)$, where $f_0$ is a
constant. As mentioned in Sec. IV A, this corresponds to the
extreme type-II situation ($\mu \gg 1$). For intermediate values
of $\mu$, one should consider both Bessel functions in Eq.
(\ref{Fint}), which is not convenient since, as mentioned in Sec.
II, this expression diverges for small V-V separation and does not
take into account neither the vortex core deformations, nor the
formation of giant vortices. Hence, using Eq. (\ref{compfit}) for
the vortex pair interaction force, with the parameters given by
Table I or by Eqs. (\ref{eqPar}a-d), would be an easy way to take
these features into account and avoid the small separation
divergence. Although this solves the problem in the standard
simulations of vortex dynamics, which involves only pairwise
interactions, here we will show when such an approach breaks down
and take the interaction between three vortices placed in the
vertices of an equilateral triangle as an example.

Let us consider three vortices placed in the positions $r_1 =
(-d/2,-\sqrt{3}d/4)$, $r_2 = (d/2,-\sqrt{3}d/4)$ and $r_3 =
(0,\sqrt{3}d/4)$, forming a triangle of side $d$. The three vortex
ansatz is $\Psi = e^{i n_1 \theta_1}e^{i n_2 \theta_2}e^{i n_3
\theta_3}f(x,y)$, where we control the vorticity $n_1$, $n_2$ and
$n_3$ of each vortex. For the present example, we consider three
singly quantized vortices $n_1 = n_2 = n_3 = 1$. Following the
procedure described in Sec. \ref{numEq}, we obtain the
Euler-Lagrange equations for the three vortex problem, where the
first equation is similar to Eq. (\ref{EL3}), but $\overline{X}$
and $\overline{Y}$ contains three terms
\begin{equation}
\overline{X} = \frac{x_1}{r_1^2}+ \frac{x_2}{r_2^2} +
\frac{x_3}{r_3^2}, \quad\quad \overline{Y} = \frac{y_1}{r_1^2}+
\frac{y_2}{r_2^2}+ \frac{y_3}{r_3^2}, \nonumber
\end{equation}
and the second equation is
\begin{equation}
\overrightarrow{\nabla} \times \overrightarrow{\nabla} \times
\overrightarrow{A} + \left[ \overrightarrow{A} -
\frac{1}{r_1}\widehat{\theta}_1 - \frac{1}{r_2} \widehat{\theta}_2
- \frac{1}{r_3} \widehat{\theta}_3 \right]f^2 = 0.
\end{equation}

\begin{figure}[!t]
\centerline{\includegraphics[width=\linewidth]{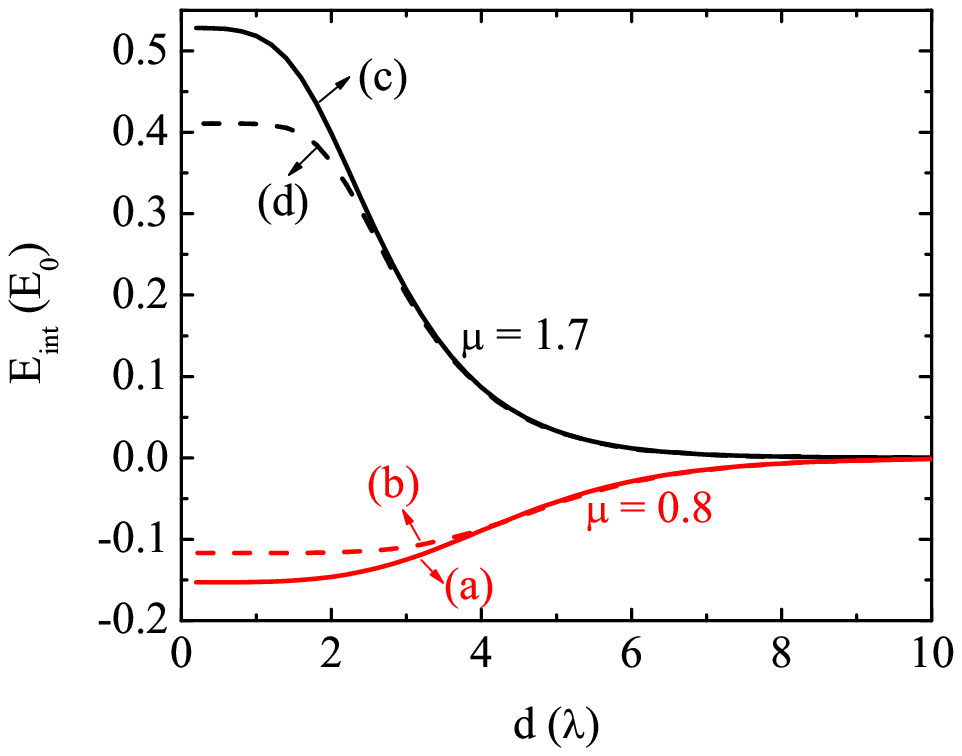}}
\caption{(Color online) Interaction energy for three vortices
placed in the vertices of an equilateral triangle as a function of
its side $d$, obtained by the three vortices ansatz (dashed) and
by considering only interaction between pairs (solid), i.e.
$E_{int}(d) = 3E_{int}^{pair}(d)$, for $\mu = 0.8$ and
1.7.}\label{fig:3V}
\end{figure}

\begin{figure}[!b]
\centerline{\includegraphics[width=\linewidth]{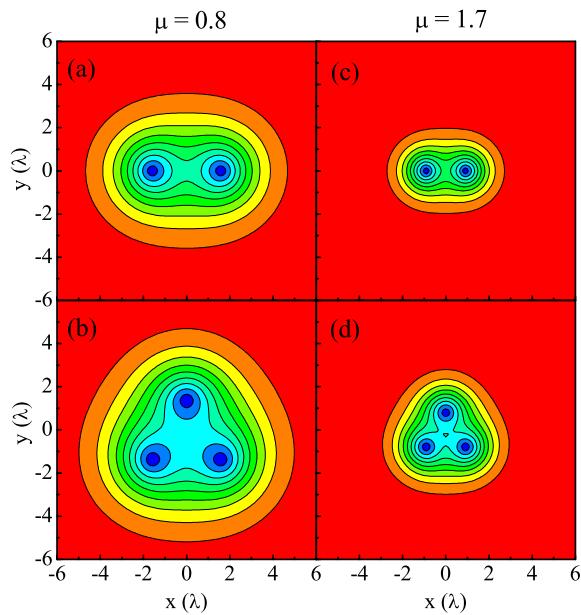}}
\caption{(Color online) Amplitude of the order parameter for the
cases indicated in Fig. \ref{fig:3V}, namely, for $\mu = 0.8$
(1.7) and $d = 3.2 \lambda$ (1.8$\lambda$), obtained for the two
(a, c) and three (b, d) interacting vortices.}\label{fig:f3V}
\end{figure}

Similarly to the case of V-V and V-GV interactions, we solved the
Euler-Lagrange equations for three vortices numerically, by means
of a finite-difference scheme and a relaxation method. The results
obtained for the interaction energy as a function of the V-V
distance, or equivalently, the side $d$ of the triangle, are shown
as dashed curves in Fig. {\ref{fig:3V}} for two values of the GL
parameter, $\mu$ = 0.8 (type-I) and 1.7 (type-II). As we consider
the same distance $d$ between each pair of vortices forming the
triangle, the standard procedure for the many-vortex problem,
which considers only pair interactions, predicts an interaction
energy $E_{int}(d) = 3E_{int}^{pair}(d)$, where
$E_{int}^{pair}(d)$ is the interaction energy for each V-V pair.
This energy is shown by the solid curves in Fig. {\ref{fig:3V}},
where good agreement with the results obtained from the three
vortex ansatz is observed only for larger separations $d$, whereas
for smaller separations the energies predicted by the pair
interaction model are clearly overestimated. This result is a
manifestation of the importance of the vortex deformations for
small V-V separation: the pairs interaction model simply does not
account for giant vortex deformations with three vortices. As a
consequence, this model overestimates the energy. This is
illustrated in Fig. \ref{fig:f3V}, where the amplitude of the
order parameter for the two (a, c) and three (b, d) interacting
vortices is shown for $\mu = 0.8$ (left panels) and 1.7 (right
panels) at V-V separations $d = 3.2\lambda$ and $1.8 \lambda$,
respectively. In the case of three vortices, we observe that each
vortex is deformed towards the center of the vortex cluster. Such
a deformation, which is found as the lowest energy state of the
three vortex system, cannot be obtained by a model consisting only
of interactions between pairs of vortices. Nonetheless, in the
extreme type-II cases studied in the literature \cite{Nori,
Pogosov}, the critical V-V separation $d_c$ where the vortices
start to coalesce approaches zero (as demonstrated in the previous
subsection) and the agreement between the results obtained by the
pairwise model and by the three vortex ansatz is expected to
improve at even smaller separations.

In the many-vortex problem, the vortices can approach each other
in a combinatorially large number of ways, and the study of three
vortices in a triangular geometry presented in this subsection is
a very specific case. Nevertheless, this example illustrates in a
simple way that, apart from the extreme type-II case, an exact
description of the many-vortex dynamics is a very difficult task.
The pair potential, even when taking V-V deformations into
account, still provides only an approximate description of the
problem, as the deformations due to the presence of all the other
vortices are not included in the model. In this sense, the
expressions proposed in the present work for the V-V and V-GV
interaction forces provide an important improvement on the
well-known expressions $\Omega(|\vec{r_i} - \vec{r_j}|) = f_0
K_1\left(|\vec{r_i} - \vec{r_j}|/\lambda\right)$ and Eq.
(\ref{Fint}), by including the deformations and the merger of
vortices, but a molecular dynamics study of many vortices using
these expressions is still not an ideal description of a system
with comparable length scales $\xi$ and $\lambda$.

\subsection{Vortex-antivortex interaction}

In the previous subsections, we showed that when two vortices or a
vortex and a giant vortex are brought close to each other, they
merge forming a single giant vortex state with vorticity $n = n_1
+ n_2$, and in the limit of small separation the V-V or V-GV
forces are very weak. Conversely, as discussed in Sec. II, a
vortex and an antivortex attract and annihilate, both in type-I
and type-II superconductors. In what follows, the behavior of the
force for the vortex-antivortex (V-AV) interaction as a function
of the V-AV separation is studied in greater detail.

\begin{figure}[!bpht]
\centerline{\includegraphics[width=\linewidth]{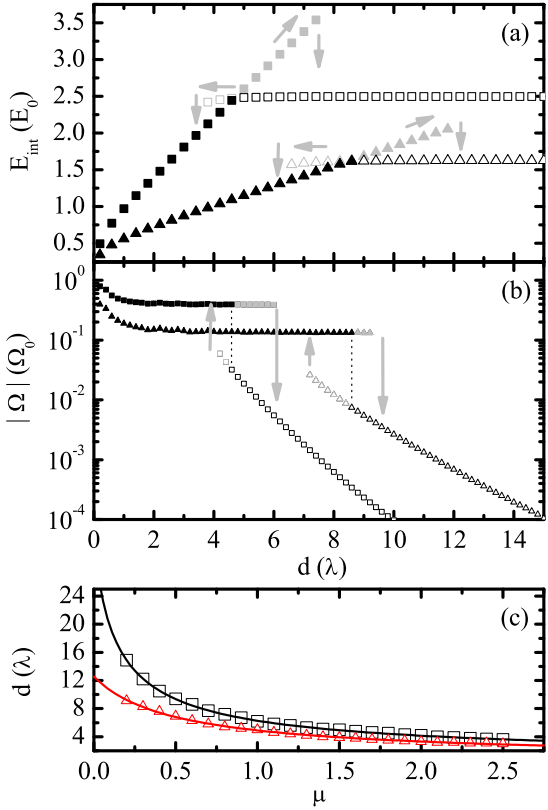}}
\caption{(Color online) V-AV interaction (a) energy and (b) force
(absolute value) as a function of the separation $d$, for $\mu =
0.6$ (triangles) and 1.7 (squares). The full (open) symbols are
the results obtained in the numerical relaxation procedure by
gradually increasing (decreasing) $d$ from 0 to 15 $\lambda$ (from
15 $\lambda$ to 0). A hysteresis is observed around a critical
separation $d_E$, as indicated by the arrows, and the solution
represented by open symbols is stable only for $d > d_A$. (c)
Numerically obtained critical separations $d_E$ (squares) and
$d_A$ (triangles) as a function of the GL parameter $\mu$, along
with their fitting functions (curve).}\label{fig:10}
\end{figure}

Indeed, the V-AV interaction is quite different from the
interaction observed in the V-V and V-GV cases studied in previous
subsections. The V-AV interaction energy (a) and force (b) are
shown in Fig. \ref{fig:10}, for two values of the GL parameter,
$\mu$ = 0.6 (triangles) and 1.7 (squares). As discussed previously
in Sec. II, the V-AV interaction is always attractive, for any
value of $\mu$. However, at some critical V-AV separation $d_E$,
the solution with well defined super-currents around each vortex
and antivortex ceases to be the lowest energy state of the system.
A solution with lower energy exhibits a strong suppression of the
amplitude of the order parameter and super-current in the region
between the vortex and the antivortex, and represents the ground
state for small V-AV distances. A hysteresis is observed in the
vicinity of the critical separation $d_E$, as shown in Fig.
\ref{fig:10}(a). These results resemble those obtained by Priour
and Fertig \cite{Priour} in the case of a vortex placed close to
an artificial defect. A suppressed amplitude of the order
parameter is also observed by Sardella \emph{et al.}
\cite{Sardella} in the dynamics of V-AV annihilation in a square
mesoscopic superconducting cylinder, for small V-AV separation.
The absolute value of the force is shown in Fig. \ref{fig:10}(b)
on $log_{10}$ scale, where two different behaviors, for
separations $d$ smaller and larger than $d_E$, are clearly
observed.

The dependence of the numerically obtained critical separation
$d_E$ for the V-AV interaction on the GL parameter $\mu$ is
illustrated as the squares in Fig. \ref{fig:10}(c), and can be
fitted to a function similar to the one used for the critical
separations in the V-V and V-GV cases, given by $d_E = 0.337 +
31.249(1+10.264\mu)^{-0.6855}$ (with estimated variance $\nu
\approx 0.4\%$), which is shown as a solid curve in Fig. 10(c).
The difference is that the $\mu \rightarrow \infty$ limit is now
finite, while previously it was zero. Although the solution with
well defined super-currents around each vortex and antivortex is
not the lowest energy state for $d < d_E$, it is still a stable
state in the vicinity of this point, and becomes unstable only for
$d < d_A$. The dependence of $d_A$ on the GL parameter $\mu$ is
shown by the triangles in Fig. \ref{fig:10}(c) and can be fitted
by $d_A = 0.337 + 12.222(1+2.461\mu)^{-0.7931}$ (with estimated
variance $\nu \approx 0.6\%$).

\begin{figure}[!b]
\centerline{\includegraphics[width=\linewidth]{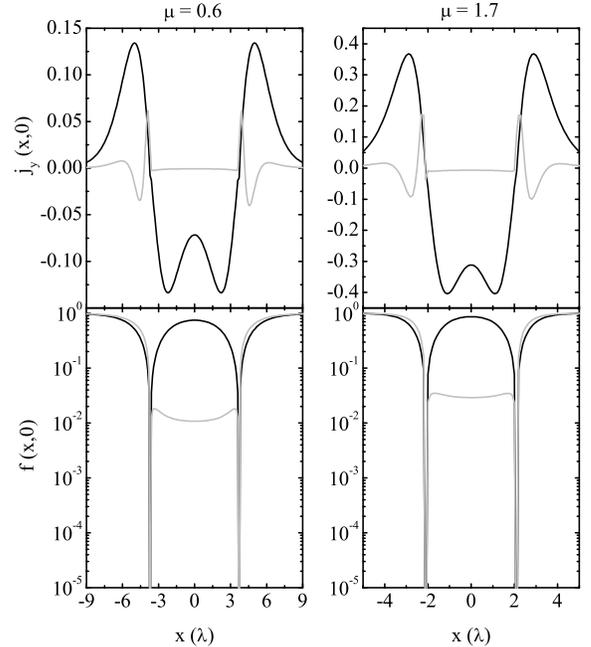}}
\caption{Super-current (top) and amplitude of the order parameter
(bottom) along the direction of the V-AV approach $x$, for the
V-AV separations indicated by arrows in Fig. 10 (b), namely, $d =
9.2 \lambda$ (5.2 $\lambda$), for $\mu = 0.6$ (1.7). Black (gray)
curves refer to the states represented by open (full) symbols in
Fig. \ref{fig:10} (b).}\label{fig:7}
\end{figure}

Figure \ref{fig:7} shows the distribution of the super-current
$\vec{J} = \vec{\nabla}\times\vec{\nabla}\times\vec{A}$ and the
amplitude of the order parameter along the direction of the V-AV
approach ($y = 0$ axis) for different values of the V-AV
separation. The critical separations for $\mu = 0.6$ and 1.7 are
$d_E = 8.6 \lambda$ and $4.5 \lambda$, respectively, and the
values of V-AV separation in Fig. \ref{fig:7} are chosen as $d =
9.2 \lambda > d_E$ for $\mu = 0.6$ and $5.2 \lambda > d_E$ for
$\mu = 1.7$. Notice that for each of these separations, we can
find two solutions with different energies. The black (gray)
curves in Fig. \ref{fig:7} are related to the open (full) symbols
in Fig. \ref{fig:10}(b). When the V-AV separation is large, the
currents around the vortex and the antivortex present well defined
peaks at some distance which depends on $\mu$. As the vortex and
antivortex are placed closer, their super-currents superimpose in
the region between them, as observed in the black curves in Fig.
\ref{fig:7} (top). The black curves in Fig. \ref{fig:7} (bottom)
show that the amplitude of the order parameter in these solutions
has zeros at each vortex and antivortex position and reaches
$\approx 1$ in the region between them. In the vicinity of $d_E$,
for V-AV separation $d > d_E$, there is a higher energy state (see
full gray symbols in Fig \ref{fig:10}(a)) with strongly suppressed
super-current and amplitude of the order parameter in the region
between the vortex and antivortex, which is shown by the gray
curves in Fig. \ref{fig:7}. For $d < d_E$, the solution
represented by black lines in Fig. \ref{fig:7} is no longer the
lowest energy state, as shown in Fig. \ref{fig:10}(a), and becomes
unstable as the V-AV separation is reduced at $d < d_A$, while the
solution with suppressed current and order parameter between the
vortex and antivortex, shown by the gray curves, becomes the
lowest energy state for $d < d_E$ and the only stable solution for
$d < d_A$.

The suppressed order parameter in the region between vortices
observed in the only stable solution for $d < d_A$ suggests that a
vortex and an antivortex cannot coexist at these distances, unless
somehow pinned, in which case this string solution is formed. This
is reasonable, since at short distances the fields of the vortex
and the antivortex compensate each other, and the flux
quantization as an essential property of a(n) (anti)vortex is
lost. Notice that this is different from the case of two merging
vortices, which can coexist at short distances, deform and
interact as described in previous sections, since the flux
quantization of the V-V pair is preserved even at small V-V
separations. The string formation goes beyond simulations of V-AV
dynamics, since in this case the vortex-antivortex pair is no
longer well defined by their surrounding super-current and order
parameter. For molecular dynamics studies of the V-AV motion, one
should consider the critical separation $d_A$ as the separation
where the V-AV pair annihilates (see, e.g. Ref. \cite{Clessio}).

\begin{table}
\caption{Fitting parameters $\Delta_i$ and estimated variance
$\nu$ for Eq. (\ref{Fdeltas}) in the V-AV case ($n_1 = 1$ and $n_2
= -1$), for $d > d_E$.} \label{tab:5}
\renewcommand{\arraystretch}{0.99}
\begin{ruledtabular}
\begin{tabular}{cccc}
  \hline
  $\mu$ & $\Delta_1$ & $\Delta_2$ & $\nu (\times10^{-9}$) \\
  \hline
 0.3 & 156.948 & 0.4203 & 0.918 \\
 0.4 & 67.315 & 0.7419 & 0.10 \\
 0.5 & 31.064 & 1.173 & 0.098 \\
 0.6 & 19.070 & 1.719 & 0.18 \\
 0.7 & 14.401 & 2.340 & 0.342 \\
 0.8 & 11.990 & 2.925 & 0.56 \\
 0.9 & 6.499 & 5.060  & 3.06\\
 1.0 & 6.357 & 6.357 & 5.01\\
 1.1 & 5.320 & 9.170 & 5.0\\
 1.2 & 4.159 & 14.269 & 1.21\\
 1.3 & 4.254 & 16.759 & 1.21\\
 1.4 & 4.125 & 20.640 & 1.10\\
 1.5 & 4.039 & 23.58 & 0.75\\
 1.6 & 3.775 & 33.662 & 1.20\\
 1.7 & 3.632 & 43.173 & 1.40\\
 1.8 & 3.542 & 51.251 & 1.03\\
 1.9 & 3.422 & 66.755 & 1.23\\
 2.0 & 3.315 & 87.448 & 1.38\\
 2.1 & 3.226 & 113.229 & 1.54\\
 2.2 & 3.162 & 138.472 & 1.25\\
 2.3 & 3.101 & 170.634 & 1.08\\
 2.4 & 3.037 & 220.086 & 1.13\\
 2.5 & 2.983 & 277.742 & 1.08\\
  \hline
\end{tabular}
\end{ruledtabular}
\end{table}

\begin{figure}[!t]
\centerline{\includegraphics[width=\linewidth]{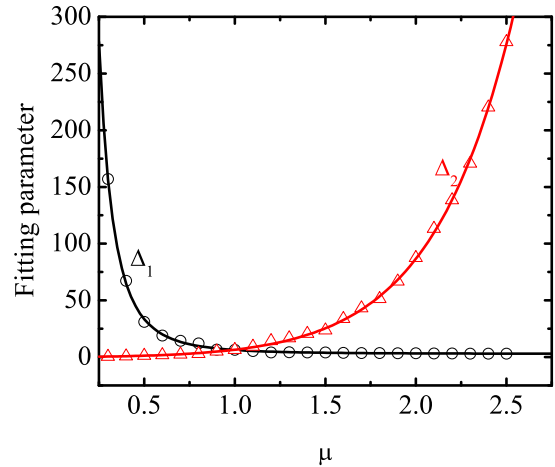}}
\caption{(Color online) Fitting functions (curves) for the
parameters $\Delta_1$ and $\Delta_2$ in Eq. (\ref{Fdeltas}), for
the V-AV interaction force at $d > d_E$, as a function of the GL
parameter $\mu = \sqrt{2}\kappa$. The data from Table \ref{tab:5}
are shown as symbols for comparison. }\label{fig:ParVAV}
\end{figure}

Due to the peculiar behavior found for the V-AV force as a
function of the separation $d$, which is discontinuous at $d_E$,
it is not possible to find a single fitting function describing
the force for both the $d > d_E$ and $d < d_E$ regimes, as we were
able to do for the V-V and V-GV forces. On the other hand, as
discussed in the previous section, the V-AV interaction force at
large distances $d$ can be described by a combination of Bessel
functions, given by Eq. (\ref{Fint}), which can be rewritten as
\begin{equation}
\Omega(d) = -\Delta_1 K_1(d) -\Delta_2 K_1(\mu d), \label{Fdeltas}
\end{equation}
where $\Delta_1$ and $\Delta_2$ are fitting parameters. We have
fitted our numerically obtained V-AV interaction force for $d >
d_E$ using Eq. (\ref{Fdeltas}), and a list of fitting parameters
for the GL parameter $\mu$ ranging from 0.3 to 2.5 is given in
Table V. A list of such fitting parameters can also be found in
Ref. \cite{Speight}, where the relation between our fitting
parameters and the parameters $q$ and $m$ of the cited work is
$\Delta_1 = m^2/2\pi^2$ and $\Delta_2 = \mu q^2/2\pi^2$. Following
the same procedure of previous subsections, we propose fitting
functions for these parameters as a function of the GL parameter
$\mu$,
\begin{subequations}\label{VAVfit}
\begin{equation}
\Delta_1 (\mu) = 2.879 + 3.415\mu^{-3.166}
\end{equation}
and
\begin{equation}
\Delta_2 (\mu) = \mu(-0.2258 + 1.044e^{1.866\mu}),
\end{equation}
\end{subequations}
which are plotted as solid lines in Fig. \ref{fig:ParVAV} along
with the data of Table \ref{tab:5} (symbols).

Notice that for the V-AV interaction, we did not find different
behaviors in different ranges of $\mu$, as observed for the V-V
and V-GV cases, since the V-AV interaction is always attractive
and becomes only stronger as $\mu$ increases from zero, instead of
exhibiting zero force at $\mu = 1$ and becoming repulsive for $\mu
> 1$, as observed for V-V and V-GV interactions. Therefore,
substituting Eqs. (\ref{VAVfit}) into Eq. (\ref{Fdeltas}) yields a
single expression
\begin{eqnarray}
\Omega(d) = -(2.879 + 3.415\mu^{-3.166}) K_1(d) \nonumber \\
+ (0.2258 - 1.044e^{1.866\mu})\mu K_1(\mu d), \label{VAVfinal}
\end{eqnarray}
which is expected to provide an accurate description of the V-AV
interaction force, at separations $d > d_E$, for any value of
$\mu$.

\section{Conclusions}

We presented a theoretical study of the interaction between
vortices in bulk superconductors within the Ginzburg-Landau
theory. An analytical study of the asymptotic behavior of the
vortex-vortex interaction shows that a combination of first order
modified Bessel functions of the second kind describes the
behavior of the numerically obtained forces for large
vortex-vortex separation. At small distances, the fitting curves
for V-V and V-GV interactions show that the force in this region
behaves as a power function of the separation between vortices. We
proposed a fitting function that combines both limiting behaviors,
namely, the power law for small distances and the modified Bessel
function behavior for large distances. This function, given by Eq.
(\ref{compfit}), gives fairly accurate fitting of the interaction
force for any value of $\mu$, even in the type-I regime. It
depends on four fitting parameters, which can be obtained for any
value of $\mu$ either by interpolating our data presented in Table
I (Table III), for vortex-vortex (vortex-giant vortex)
interactions, or by using Eqs. (\ref{eqPar} a-d) with the
parameters presented in Table II (Table IV).

Our analytical study of the V-AV interaction shows that the V-AV
interaction force is attractive for any value of the GL parameter
$\mu$, which is confirmed by our numerical results and contradicts
the conjecture proposed in previous works \cite{Moshchalkov1,
Moshchalkov2, Teniers} which implies that the V-AV interaction
force is repulsive for $\mu < 1$ (type-I). For large V-AV
separation $d$, the interaction force decays with $d$ as a
combination of modified Bessel functions. However, for $d$ smaller
than a critical separation $d_E$, the conventional V-AV pair is no
longer the lowest energy state. Instead, the lowest energy state
exhibits a strong suppression of the super-current and amplitude
of the order parameter in the region between the vortex and
antivortex, which results in a different behavior of the force as
a function of $d$ in this case and, as a consequence, the V-AV
interaction force is discontinuous at $d_E$. Furthermore, the
conventional V-AV pair becomes unstable for $d$ lower than the
separation $d_A$, which is interpreted as the V-AV annihilation
point. We fitted the interaction force for V-AV separations $d
> d_E$ by Eq. (\ref{Fdeltas}) and proposed an approximate
analytical expression for the V-AV interaction force at these
separations, given by Eq. (\ref{VAVfinal}), which is valid for any
value of $\mu$.

The fitting functions for the V-V, V-GV and V-AV force given in
this work will be useful, for instance, for the study of bulk and
mesoscopic systems consisting of many vortices using molecular
dynamics techniques. We nevertheless remark that, although
deformations are taken into account in the interaction force
between two vortices in this work, the deformations in a many
vortices system are expected to be more complex. Hence, the
molecular dynamics study of many vortices, even with the improved
expressions for the interaction force provided in this paper, is
still an approximate description of the system. As a method, the
derivation and handling of the differential equations describing
the interaction between vortices presented in this work can be
further adapted to describe such interactions in e. g. two-band
superconductors, or hybrid systems comprising different
superconducting materials.

\acknowledgements This work was financially supported by CNPq,
under contract NanoBioEstruturas 555183/2005-0, FUNCAP, CAPES, the
Bilateral programme between Flanders and Brazil, the collaborative
project CNPq-FWO-Vl, the Belgian Science Policy (IAP) and the
Flemish Science Foundation (FWO-Vl).

\newpage

\end{document}